\documentclass[universe,article,accept,pdftex,moreauthors]{Definitions/mdpi} 

\firstpage{1} 
\pubvolume{1}
\issuenum{1}
\articlenumber{0}
\pubyear{2025}
\copyrightyear{2025}
\externaleditor{Firstname Lastname} 
\datereceived{23 April 2025 } 
\daterevised{7 May 2025 } 
\dateaccepted{9 May 2025 } 
\datepublished{ } 
\hreflink{https://doi.org/} 

\newcommand{\aap}{Astron. Astrophys.}
\newcommand{\aj}{Astron. J.}
\newcommand{\apj}{Astrophys. J.}
\newcommand{\apjs}{Astrophys. J. Suppl. Ser.}
\newcommand{\apjl}{Astrophys. J. Lett.}
\newcommand{\mnras}{Mon. Not. R. Astron. Soc.}
\newcommand{\aapr}{Astron. Astrophys. Rev.}
\newcommand{\aaps}{Astron. Astrophys. Suppl.}
\newcommand{\pasp}{Publ. Astron. Soc. Pac.}

\newcommand{\araa}{Annu. Rev. Astron. Astrophys.}

\Title{Superluminal Motion and Jet Parameters in the High-Redshift Blazar J1429+5406 
}

\TitleCitation{Superluminal Motion and Jet Parameters in the High-Redshift Blazar J1429+5406}

\Author{D\'avid Koller $^{1,2,3}$\orcidA{} and S\'andor Frey $^{2,3,4,}$*\orcidB{}}

\AuthorNames{D\'avid Koller and S\'andor Frey}

\AuthorCitation{Koller, D.; Frey, S.}

\address{%
$^{1}$ \quad Department of Astronomy, Institute of Physics and Astronomy, ELTE E\"otv\"os Lor\'and University,  \linebreak P\'azm\'any P\'eter s\'et\'any 1/A, H-1117 Budapest, Hungary; koller.david@csfk.org 
\\
$^{2}$ \quad Konkoly Observatory, HUN-REN Research Centre for Astronomy and Earth Sciences,  \linebreak Konkoly Thege Mikl\'os \'ut 15-17, H-1121 Budapest, Hungary\\
$^{3}$ \quad CSFK, MTA Centre of Excellence, Konkoly Thege Mikl\'os \'ut 15-17, H-1121 Budapest, Hungary\\
$^{4}$ \quad Institute of Physics and Astronomy, ELTE E\"otv\"os Lor\'and University, P\'azm\'any P\'eter s\'et\'any 1/A, H-1117~Budapest, Hungary
}

\corres{Correspondence: frey.sandor@csfk.org}

\abstract{We investigate the relativistic jet of the powerful radio-emitting blazar J1429+5406 at redshift $z=3.015$. Our understanding of jet kinematics in $z \geq 3$ quasars is still rather limited, based on a sample of less than about $50$ objects. The blazar J1429+5406 was observed at a high angular resolution using the method of very long baseline interferometry over more than two decades, between 1994 and 2018. These observations were conducted at five radio frequencies, covering a wide range from $1.7$ to $15$~GHz. The outer jet components at $\sim$20--40~milliarcsecond (mas) separations from the core do not show discernible apparent motion. On the other hand, three jet components within the central $10$ mas region exhibit significant proper motion in the range of (0.045--0.16)~mas\,year$^{-1}$, including one that is among the fastest-moving jet components at $z \geq 3$ known to date. Based on the proper motion of the innermost jet component and the measured brightness temperature of the core, we estimated the Doppler factor, the bulk Lorentz factor, and the  inclination angle of the jet with respect to the line of sight. The core brightness temperature is at least $3.6 \times 10^{11}$\,K, well exceeding the equipartition limit, indicating Doppler-boosted radio emission. The low jet inclination ($\lesssim$5.4$^{\circ}$) firmly places J1429+5406 into the blazar category.} 

\keyword{high-redshift galaxies; active galactic nuclei; blazars; relativistic jets; radio continuum; very long baseline interferometry}

\begin{document}


\section{Introduction}

The largely unexplored, high-redshift ($z\gtrsim3$) domain of the Universe is key to understanding the evolution of active galactic nuclei (AGN) \cite{2020ApJ...897..177P}. By~analysing these objects, we can determine whether AGN in the early Universe functioned similarly to those in our local cosmic~neighbourhood.

Supermassive black holes (SMBHs) are an essential component of AGN. These black holes have masses ranging from approximately $10^{6}$ to $10^{10}$ solar masses~\cite{2017A&ARv..25....2P} and are believed to reside at the center of every major galaxy~\cite{2013ARA&A..51..511K}. As~matter spirals into the SMBH, the~galactic core can reach extremely high bolometric luminosities ($L_\mathrm{bol}$$\sim$$10^{48}$~erg\,s$^{-1}$), thousands of times brighter than an entire normal galaxy~\cite{2017A&ARv..25....2P}. Under~the influence of the strong magnetic field, not all of the accreted plasma falls into the SMBH; instead, some material is ejected along the SMBH spin axis at relativistic speeds~\cite{1982MNRAS.199..883B}. The~magnetic field collimates this outflow into bipolar relativistic jets~\cite{1982MNRAS.199..883B}. The~charged particles experience acceleration along helical trajectories due to the Lorentz force in the magnetic field. As~a result, they emit synchrotron radiation that can be detected in the radio~\cite{1984RvMP...56..255B}. The~reason why only $\lesssim$$10\%$ \cite{2002AJ....124.2364I} of known AGN are radio-loud and ``jetted'' is still an open question~\cite{2019A&A...625A..25M,2024MNRAS.532.3036M}.

The technique of very long baseline interferometry (VLBI) \cite{2022Univ....8..527J} uses coordinated observations performed by an extended network of radio telescopes to~achieve the highest possible angular resolution. With~a global VLBI array, resolutions of around a milliarcsecond (mas) can be achieved at cm wavelengths. In~many AGN, instead of symmetric two-sided jets, strongly asymmetric structures are observed with VLBI. The~appearance of one-sided core--jet structures is attributed to Doppler beaming, a~relativistic effect where the emission from the approaching jet closely aligned with the observer's line of sight is boosted, while the emission from the opposite (receding) jet is diminished~\cite{1995PASP..107..803U}. 

In blazars, the~jet is viewed at an angle of 10° or less~\cite{2022Galax..10...35P}, causing its Doppler-boosted emission to dominate the spectral energy distribution. Blazars are further categorized into two main types: BL Lac objects and flat-spectrum radio quasars (FSRQs). FSRQs exhibit strong and broad optical emission lines, whereas BL Lac objects have weak or no optical emission lines as they are suppressed by the continuum. Additionally, FSRQs possess higher bolometric luminosities~\cite{1997ApJ...487..536S}. As~a result, they dominate the known high-redshift ($z\gtrsim3$) blazar population, as~they are significantly brighter than BL Lac~objects.

Some studies suggest that SMBHs grow faster in jetted quasars compared to radio-quiet AGN~\cite{2015MNRAS.446.2483S}, indicating a connection between jet activity and black hole growth~\cite{2014MNRAS.442L..81F}. The~most distant known blazar has a redshift of nearly $z=7$ \cite{2024ApJ...977L..46B}, while AGN have been discovered with redshifts as high as $z=10$ \cite{2024NatAs...8..126B}. However, jetted AGN do not appear to be dominant in the early Universe, as~they have only been observed up to $z<7$ \cite{2021ApJ...909...80B}. Radio-loud quasars are rare even beyond $z\approx 4$, based on current observations~\cite{2017FrASS...4....9P}.

According to a recent work~\cite{2025Univ...11...91G}, the~number of radio quasars at $z\ge3$ whose jet kinematics and physics have been studied by means of multi-epoch high-resolution VLBI observations is currently about $50$ \cite{2010A&A...521A...6V,2015MNRAS.446.2921F,2018MNRAS.477.1065P,2020SciBu..65..525Z,2020NatCo..11..143A,2022ApJ...937...19Z,2024Univ...10...97F,2024A&A...689A..43B,2024MNRAS.530.4614K}.
The VLBI jet kinematic analysis of the flat-spectrum radio quasar~\cite{2007ApJS..171...61H} J1429+5406 (TXS~1427+543) presented here provides a valuable addition to the sample studied to date, offering an interesting comparison with other high-redshift jetted AGN. The~right ascension and declination coordinates of the object in the VLBI-based International Celestial Reference Frame~\cite{2020A&A...644A.159C} are RA $=14^\mathrm{h}29^\mathrm{min}21.87879^\mathrm{s}$ and DEC$= 54^{\circ}06^\prime 11.1228^{\prime\prime}$, respectively, at 2015.0 as the reference epoch. We consider the spectroscopic redshift determined recently as $z=3.0151 \pm 0.0002$ based on measurements from the Dark Energy Spectroscopic Instrument~\cite{2024AJ....168...58D}. The~radio source has a two-point spectral index of $\alpha_\mathrm{1.4\,GHz}^\mathrm{3.0\,GHz}=-0.68$ \cite{2024MNRAS.530.2590H} (following the convention $S \propto \nu^\alpha$, where $S$ is the flux density and $\nu$ the observing frequency). The~flux densities of J1429+5406 are $1165$~mJy at $1.4$~GHz in the FIRST (Faint Images of the Radio Sky at Twenty-centimeters) survey~\cite{1997ApJ...475..479W}, $792$~mJy at $3$~GHz in the VLASS (Very Large Array Sky Survey) \cite{2020RNAAS...4..175G}, and~$493$~mJy at $8.4$~GHz measured with the VLA~\cite{1992MNRAS.254..655P}. While the source is compact on arcsecond scales~\cite{1997ApJ...475..479W,2020RNAAS...4..175G}, \mbox{$8.4$-GHz}
VLA imaging at $\sim$$0.2^{\prime\prime}$ resolution revealed an extension in the $110^{\circ}$ position angle~\cite{1992MNRAS.254..655P}.
The source was also detected in X-rays by \textit{ROSAT} with the flux of $(3.6\pm 1.1) \times 10^{-13}$~erg\,s$^{-1}$\,cm$^{-2}$ in the (0.1--2.4)~keV band~\cite{2007A&A...476..759B} and~in $\gamma$-rays by the \textit{Fermi } Large Area Telescope~\cite{2020ApJ...892..105A,2020ApJ...897..177P}.
It is faint in the optical, with~apparent magnitude $m_\mathrm{V}\simeq20.4$~\cite{2010A&A...518A..10V}.
 
Here we present the analysis of archival multi-frequency VLBI observations of J1429+5406 obtained between 1994 and 2018. Section~\ref{sec:Observations and Data Analysis} details the observations and data reduction, Section~\ref{sec:Results} presents the results, Section~\ref{sec:Discussion} provides a discussion, and~Section~\ref{sec:Summary and Conclusions} summarizes our~findings.

Throughout this paper, we assume a standard flat $\Lambda$ Cold Dark Matter cosmological model with parameters \( H_0 = 67.8 \,\mathrm{km\,s^{-1}\,Mpc^{-1}} \), \( \Omega_\mathrm{m} = 0.308 \), and~\( \Omega_\Lambda = 0.692 \) \cite{2016A&A...594A..13P}. In~this model, the~luminosity distance of J1429+5406 is $D_\mathrm{L} = 26196$~Mpc and~$1$~mas angular size corresponds to $7.878$~pc projected linear size at $z=3.015$ \cite{2006PASP..118.1711W}. 

\section{Observations and Data~Analysis}
\label{sec:Observations and Data Analysis}{}

Most of the datasets analyzed here were observed with the Very Long Baseline Array (VLBA). This network consists of ten 25 m diameter antennas in the United States: Brewster (BR), Fort Davis (FD), Hancock (HN), Kitt Peak (KP), Los Alamos (LA), North Liberty (NL), Owens Valley (OV), Pie Town (PT), Mauna Kea (MK), and~St. Croix (SC). These calibrated visibility data were obtained from the Astrogeo database (\href{https://astrogeo.org/cgi-bin/imdb_get_source.csh?source=J1429\%2B5406}{https://astrogeo.org/cgi-bin/imdb\_get\_source.csh?source=J1429\%2B5406}, accessed on 18 April 2025). In~addition, two datasets were obtained using the European VLBI Network (EVN). The~following radio telescopes were involved in at least one EVN observation: 
Effelsberg (EF; Germany), Jodrell Bank Lovell and Mk2 telescopes (JB and JV; United Kingdom), Medicina (MC; Italy), Toru\'n (TR; Poland), Onsala (ON; Sweden), Sheshan and Nanshan (SH and UR; China), Badary and Zelenchukskaya (BD and ZC; Russia), and~the phased array of the Westerbork Synthesis Radio Telescope (WB; The Netherlands). The~EVN data were obtained in the project EF022 (PI: S. Frey) in 2010, when J1429+5406 was used as a phase-reference calibrator for the nearby faint radio quasar J1429+5447 at an extremely high redshift ($z = 6.21$) \cite{2011A&A...531L...5F}. Altogether, the~five observed frequency bands were $1.7$, $2.3$, $5$, 8.3--8.7, and~$15.4$~GHz (L, S, C, X, and~U bands, respectively). For~the sake of simplicity, we collectively refer to the frequency of X-band observations as $8$~GHz hereafter. The~observing parameters for each dataset are collected in Table~\ref{tab:obs}, where the participating antennas are listed by their two-letter station codes. The~minus sign preceding a VLBA station code indicates a telescope not included in the~network. 

The visibility data downloaded from the Astrogeo archive have previously undergone initial amplitude and phase calibration, as~well as the EVN data obtained in the experiment EF022~\cite{2011A&A...531L...5F}. We imported these data to the  \textsc{Difmap} program~\cite{1997ASPC..125...77S} for imaging and model fitting. To~produce the VLBI images of J1429+5406, we used conventional hybrid mapping with cycles of \textsc{clean} deconvolution~\cite{1974A&AS...15..417H} and self-calibration~\cite{1984ARA&A..22...97P}. Finally, to~quantitatively characterize the brightness distribution, we fitted circular two-dimensional Gaussian model components~\cite{1995ASPC...82..267P} directly to the self-calibrated visibility~data.
\vspace{-4pt}
\begin{table}[H]
\caption{Details of VLBI observations of J1429+5406.\label{tab:obs}}
	\begin{adjustwidth}{-\extralength}{0cm}

\begin{tabular}{m{2cm}<{\centering}m{1cm}<{\centering}m{8cm}<{\centering}m{0.93cm}<{\centering}m{1.5cm}<{\centering}m{2.5cm}<{\centering}}

			\toprule
			\textbf{Epoch}	& \boldmath{$\nu$}	& \textbf{Stations} &
            \boldmath{$t$}&
            
            \textbf{IF} \boldmath{$\times$} \textbf{BW} &
            \textbf{Project Code}\\

\textbf{(year)}	& \textbf{(GHz)}	& \textbf{}&
\textbf{(s)}&

\textbf{(MHz)}&
\textbf{}\\
			\midrule
\multirow[m]{2}{*}{1994.612 *}	& 2.27 	&		\multirow{2}{*}{VLBA}		& \multirow{2}{*}{177}&  \multirow{2}{*}{$4\times4$}& \multirow{2}{*}{BB023~\cite{2002ApJS..141...13B}}\\\
			  	                   & 8.34		
                                    &
                                   
                                   & 
                                   &
                                   
                                   &
                                   \\
			             	     
\midrule
1998.122 *& 5.00 & VLBA & 2416 & $2\times8$ & {BV025A~\cite{2007A&A...472..763B,2007ApJ...658..203H}} \\
\midrule
2010.149 *& 15.37 & VLBA (–HN) & 5178 & $4\times8$ & BG197B \\
\midrule
2010.401 & 4.99 & EF, JB, JV, MC, TR, ON, SH, UR, BD, ZC, WB & 21,105 &  $8 \times 16$ & {EF022A~\cite{2011A&A...531L...5F}}\\
\midrule
2010.434 & 1.66 & EF, JB, MC, TR, ON, SH, UR, BD, ZC, WB & 21,102 & $8 \times 16$ & {EF022B~\cite{2011A&A...531L...5F}}\\
\midrule
\multirow[m]{2}{*}{2017.235 *}	& 2.29 	&		\multirow{2}{*}{VLBA}		& \multirow{2}{*}{339}& $4\times32$& \multirow{2}{*}{UF001F~\cite{2021AJ....162..121H}}\\\
			  	                   & 8.67		
                                    &
                                   
                                   &
                                
                                   & $12\times 32$
                                   &
                                   \\
    \midrule
\multirow[m]{2}{*}{2018.346 *}	& 2.25 	&		\multirow{2}{*}{VLBA (–OV)}		& \multirow{2}{*}{288}& $3\times32$& \multirow{2}{*}{UG002G}\\\
			  	                   & 8.65		
                                    &
                                   
                                   & 
                                   
                                   & $12\times 32$
                                   &
                                   \\

			\bottomrule
		\end{tabular}
	\end{adjustwidth}
\noindent{\footnotesize{Notes: * data obtained from the Astrogeo database. Col.~1: mean epoch in year; Col.~2: central observing frequency; Col.~3: antennas involved in the measurements; Col.~4: on-source integration time; Col.~5: number of intermediate frequency channels (IF) times the bandwidth per IF; Col.~6: VLBI project code and literature reference (if available).}}
\end{table}
\unskip

\section{Results}
\label{sec:Results}
\unskip

\subsection{Core--Jet Structure at Multiple~Frequencies}

\begin{figure}[H]

\includegraphics[width=12 cm]{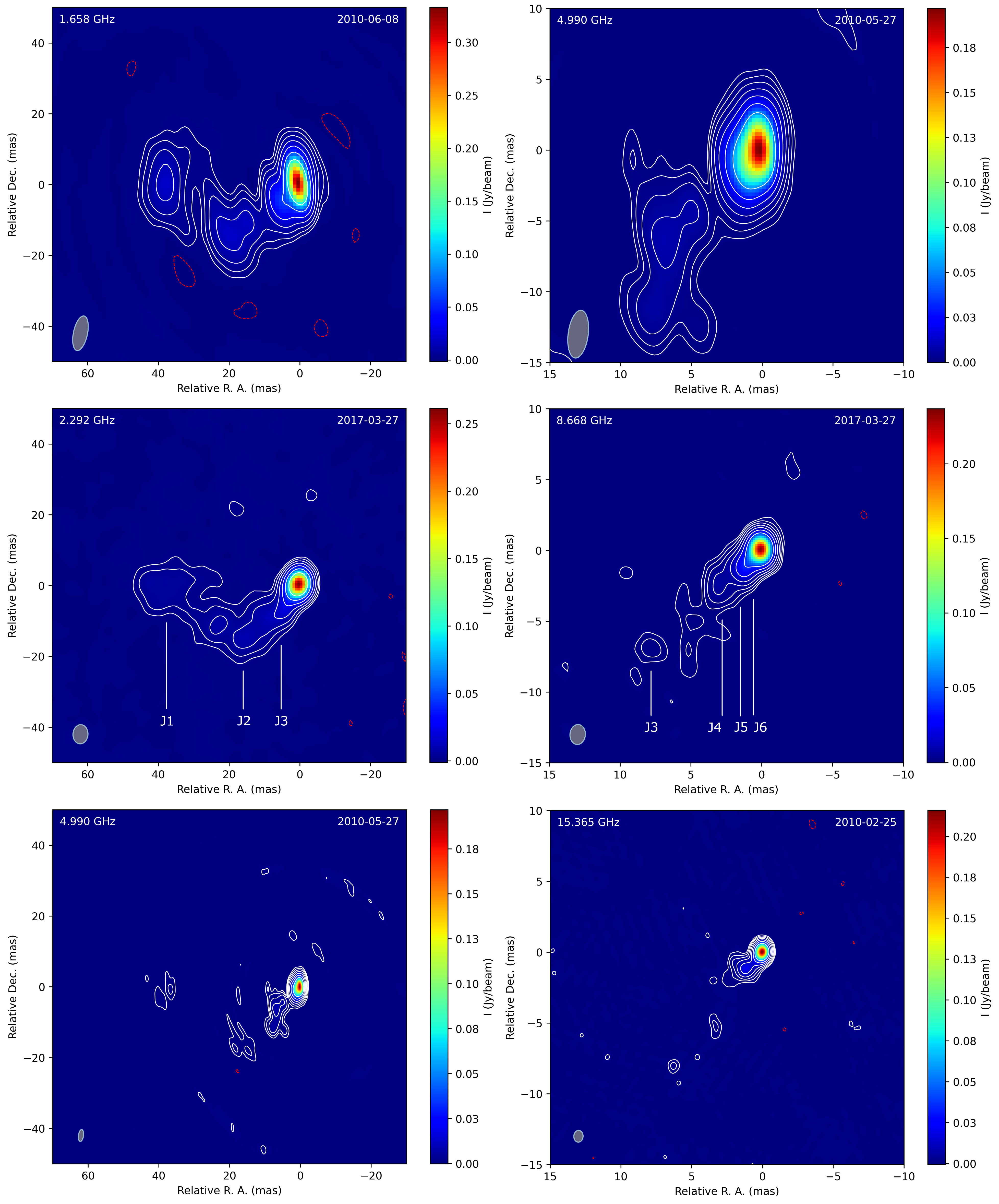}

\caption{Selected naturally weighted VLBI images of J1426+5406 at five different frequencies ($1.7$, $2.3$, $5$, $8$, and~$15.4$~GHz) from between the years 2010 and 2017. The~images are centred on their brightness  \label{fig:images}}
\end{figure}

Figure~\ref{fig:images} shows example VLBI images of J1429+5406 at each of the frequencies used for observations. The~maps presented are based on observations between 2010 and 2017 and are arranged from the lowest to the highest frequency. The~maps clearly show the compact, bright core along with multiple extended and fainter jet components at various separations from the core. The~image parameters are summarized in Table~\ref{tab:image}.

\begin{figure}[H]

\includegraphics[width=11 cm]{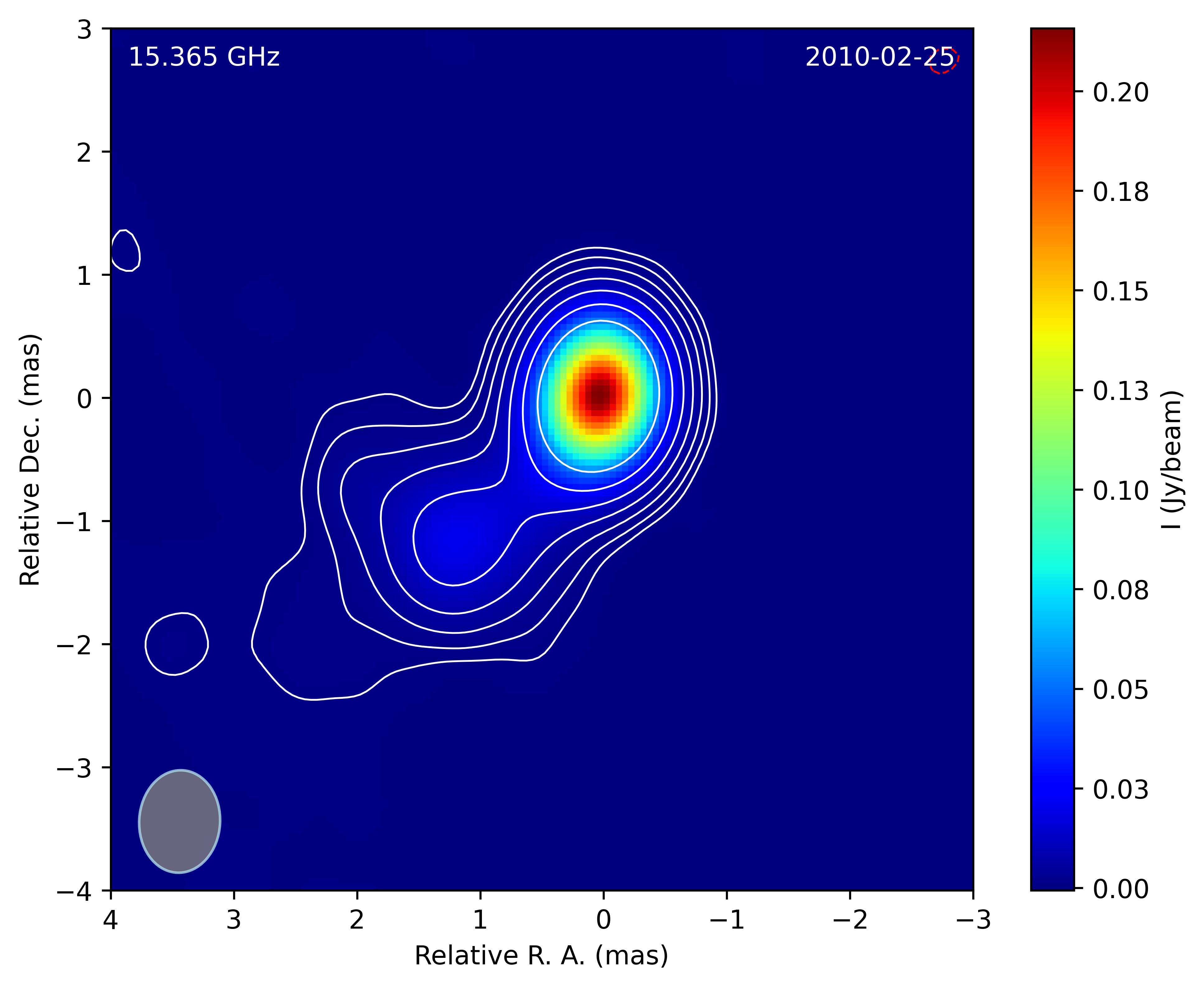}
\caption{The innermost $\sim$$3$ mas section of the core--jet structure of J1426+5406 in the central part of the $15.4$ GHz image shown in Figure~\ref{fig:images}. See Figure~\ref{fig:images}'s caption for a detailed description, Table~\ref{tab:image} for the image parameters, and Table~\ref{tab_U} for the positions of the fitted jet~components. \label{fig:uband}}
\end{figure}

The images within the left and right columns in Figure~\ref{fig:images} display the same fields of view to~facilitate easy comparison between the different frequencies. 
While the $1.7$, $2.3$, and~$5$ GHz images on the left of Figure~\ref{fig:images} illustrate the full extent of the jet structure revealed by these VLBI observations, the~images in the right column zoom into the inner section of the jet at $5$, $8$, and~$15.4$~GHz (the $5$ GHz image is seen in both the left and right columns, but~with different relative right ascension and declination ranges.) Finally, the~innermost core--jet region is shown in Figure~\ref{fig:uband} where the central part of the $15.4$ GHz image is reproduced. The diminishing outer extended jet features at higher frequencies are generally consistent with the decreasing spectral index at increasing distances from the core due to spectral aging~\cite{2012A&A...544A..34P}, as~well as the increasing angular resolution.
\vspace{-4pt}
   {\captionof*{figure}{peak. The~intensities are represented by colors according to the palettes on the right-hand side of each panel, as~well as the contours whose first level is drawn at approximately $\pm 3\sigma$ image root mean square (rms) noise, except~for the $5$ GHz image ($\sim$$4\sigma$). The~positive contours increase by a factor of $2$. Negative contours are shown as red dashed curves. The~parameters of the images and the elliptical Gaussian restoring beams are given in Table~\ref{tab:image}. The~half-power width of the restoring beam is illustrated by the ellipse in the lower-left corner of each image. The~approximate positions of the fitted jet components (see Tables~\ref{tab_L}--\ref{tab_U}) are indicated in the middle~panels.}}

\unskip   

\begin{table}[H] 
\caption{Parameters of the VLBI images shown in Figures~\ref{fig:images} and \ref{fig:uband}.\label{tab:image}}

\begin{tabularx}{\textwidth}{CCCCCCC}
\toprule
\boldmath{$\nu$}	&
\boldmath{$I_\mathrm{p}$}	&  \boldmath{$I_0$} &
\textbf{rms} &
\boldmath{$\theta_\mathrm{maj}$} &
\boldmath{$\theta_\mathrm{min}$} &
\textbf{PA} 
\\

\textbf{(GHz)}& \multicolumn{3}{c}{\textbf{(mJy\,beam}\boldmath{$^{-1}$}\textbf{)}} &
\multicolumn{2}{c}{\textbf{(mas)}} &
\textbf{(}\boldmath{$^\circ$}\textbf{)}\\
\midrule
1.66 		& 	333		&$\pm1.9$ & 0.62&9.96& 3.99 & 11.9\\
2.29 		& 	261		&$\pm1.7$ & 0.58&5.43& 4.24 & 3.7\\
4.99		& 	197		&$\pm0.80$ & 0.18&3.40& 1.42 & 7.5\\
8.67		& 	237		&$\pm0.61$ &0.20 &1.41& 1.09 & 5.0\\
15.37 		& 	216		&$\pm0.84$ & 0.28&0.83& 0.66 & 3.2\\
\bottomrule
\end{tabularx}

\noindent{\footnotesize{Notes: Col.~1: central observing frequency; Col.~2: peak intensity; Col.~3: lowest intensity contour level; C\mbox{ol.~4: rm}s noise level in the residual map; Col.~5: elliptical Gaussian restoring beam major axis (half-power width); C\mbox{ol.~6: r}estoring beam minor axis (half-power width); Col.~7: restoring beam major axis position angle, measured from north through east.}}
\end{table}

\vspace{-14pt}

\subsection{The Outer Jet Components and Their Proper~Motions}

The VLBI maps indicate that the jet undergoes a directional change. At~the lower frequencies ($1.7$ and $2.3$~GHz), which are sensitive to extended structures (Figure~\ref{fig:images}), the~jet's projection onto the plane of the sky shows a significant turn at approximately $20$~mas from the core, changing from southeast to northeast. According to the fitted model parameters (Tables~\ref{tab_L} and \ref{tab_S}), the~inner jet components are located at position angles $\sim$$140^{\circ}$, while the outermost J1 component has a position angle of $\sim$$90^{\circ}$, indicating a change in the position angle of $\sim$$50^{\circ}$. Note that the position angle of the reported sub-arcsecond extension~\cite{1992MNRAS.254..655P}, $110^{\circ}$, is broadly consistent with the VLBI jet direction, although~the $8.4$ GHz VLA image of J1429+5406 is not published~\cite{1992MNRAS.254..655P}. 

\begin{figure}[H]
\includegraphics[width=13.8 cm]{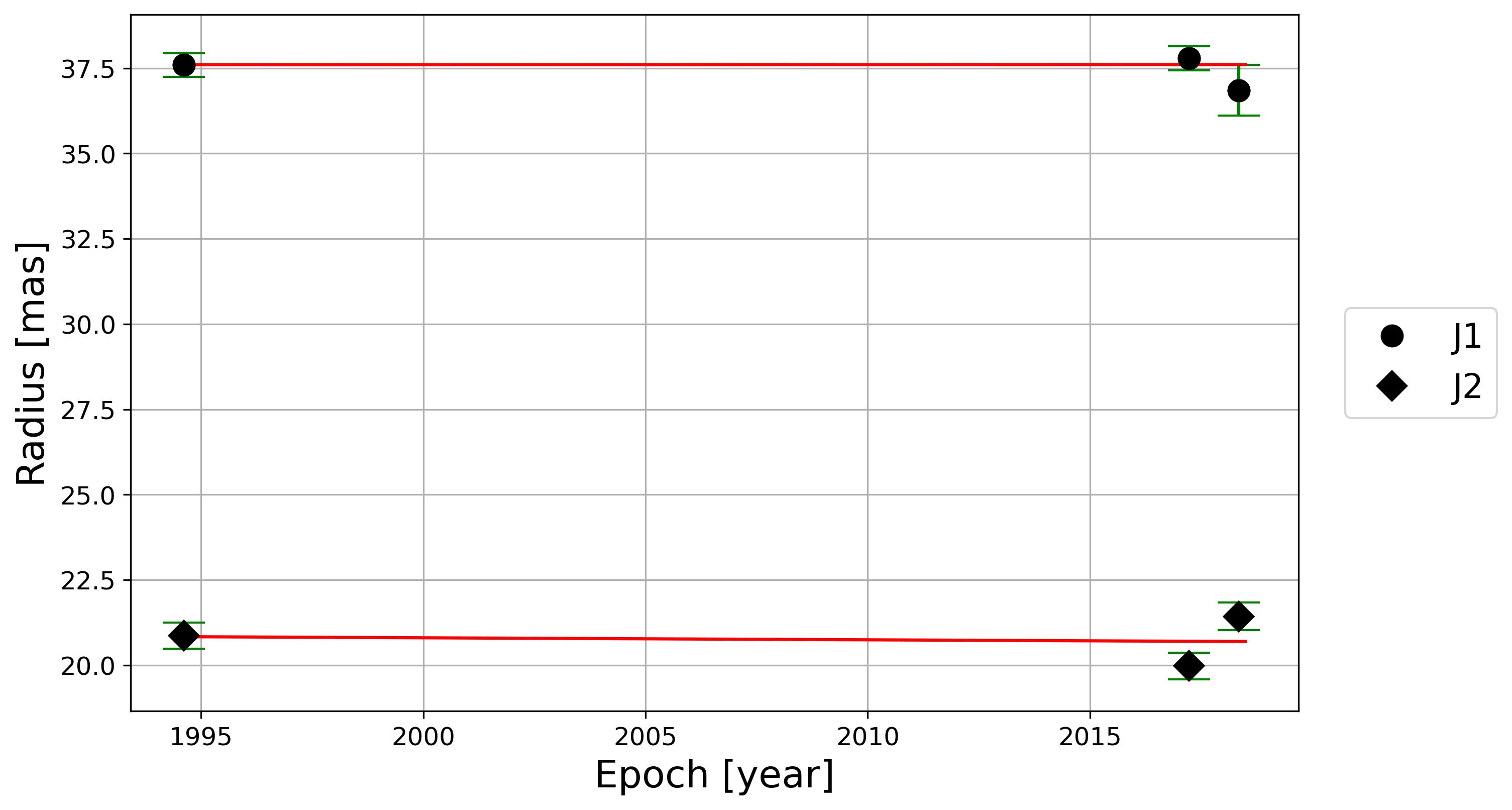}
\caption{Core--jet distances as a function of time in J1429+5406, based on $3$ epochs of VLBI observations and brightness distribution model fitting at $2.3$~GHz (S band). The~slopes of the fitted lines do not indicate significant apparent jet component proper motions for J1 and J2.\label{fig:propermotion_L}}
\end{figure}

We calculated the apparent proper motions of the jet components with respect to the core at $2.3$~GHz. There is no significant change in the position angle and the separation from the core (Figure~\ref{fig:propermotion_L}) for the J1 and J2 components over the period covered by the VLBI observations (1994--2018).
For the J3 component, however, proper motion was detected at $2.3$~GHz.
The fitted slopes and their formal uncertainties are 
\mbox{$\psi_\mathrm{J1}=(-0.0003\pm0.02)$~mas\,year$^{-1}$,}
$\psi_\mathrm{J2}=(-0.01\pm0.05)$~mas\,year$^{-1}$, 
and \mbox{$\psi_\mathrm{J3}=(0.11\pm0.02)$~mas\,year$^{-1}$.}

\unskip   

\subsection{Proper Motions of the Inner Jet~Components}

\begin{figure}[H]
\includegraphics[width=13.8 cm]{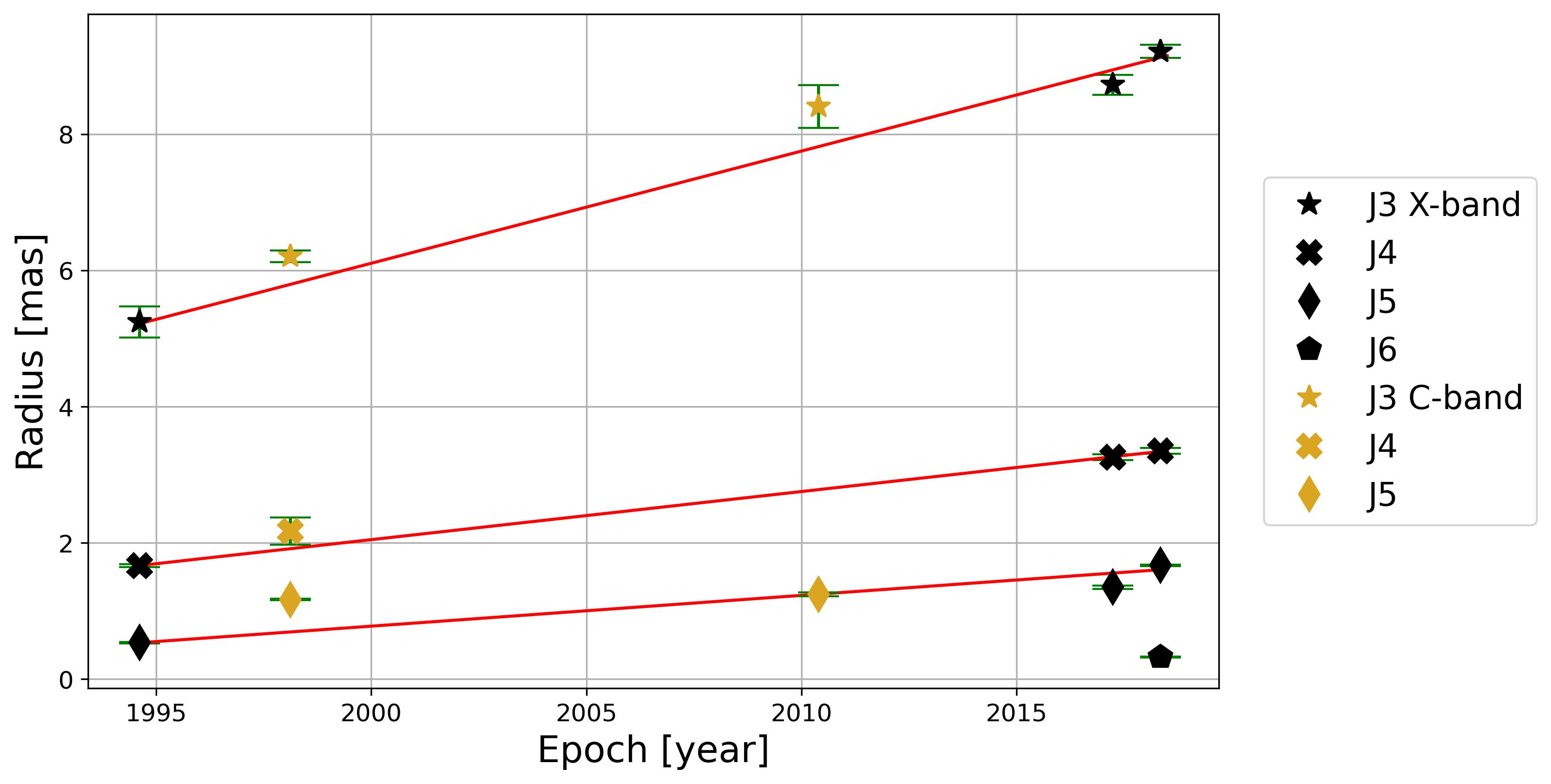}
\caption{Core--jet distances as a function of time in J1429+5406, based on $3$ epochs of VLBI observations and brightness distribution model fitting at $8$~GHz (X band). The~slopes of the fitted lines indicate apparent jet component proper motions determined for J3, J4, and~J5. The~presumably newly emerging component J6 was identified at the last epoch only. The~$5$ GHz (C-band) data points are shown in yellow for illustration purposes only and~were not used for the proper motion~determination.\label{fig:propermotion}}
\end{figure}

Based on the circular Gaussian brightness distribution model components fitted to the self-calibrated visibility data at $8$~GHz, we determined the core--jet distances ($R$) at the three~epochs available (Table~\ref{tab_X}). From~these data, we determined the apparent proper motion of each jet component that could be identified at this frequency band. We obtained \mbox{$\mu_\mathrm{J3}=(0.16\pm0.02)$~mas\,year$^{-1}$}, $\mu_\mathrm{J4}=(0.0706\pm0.0004)$~mas\,year$^{-1}$, and $\mu_\mathrm{J5}=(0.045\pm0.006)$~mas\,year$^{-1}$ for components J3, J4, and~J5, respectively, by~fitting linear functions, as shown in Figure~\ref{fig:propermotion} (note that the values of $\psi_\mathrm{J3}$ and $\mu_\mathrm{J3}$ obtained at $2.3$ and $8$~GHz, respectively, are close to each other). To illustrate the general consistency of the proper motion estimates, in~Figure~\ref{fig:propermotion} we also indicate the data points measured at a close frequency ($5$~GHz, Table~\ref{tab_C}) that  were not included in the linear~regression. 

The apparent speeds of the jet components can be calculated in the units of the speed of light ($c$) based on the following equation~\cite{2012ApJ...760...77A}:
\begin{linenomath}
\begin{equation}
\beta_{\text{app}} = 0.0158 \times \frac{\mu D_\text{L}}{1 + z},
\end{equation}
\end{linenomath}
where the apparent proper motion $\mu$ is measured in mas\,year$^{-1}$. For~components with detectable proper motion at $8$~GHz, these values are $\beta_\mathrm{J3}=(17.0\pm1.9)\,c$, \mbox{$\beta_\mathrm{J4}=(7.28\pm0.04)\,c$}, and~$\beta_\mathrm{J5}=(4.7\pm0.6)\,c$. Each calculated $\beta_\mathrm{app}$ value exceeds unity, indicating that these components exhibit apparent superluminal~motion.

\subsection{Inner Jet~Parameters}

Based on the $8$ GHz measurements, we determined the brightness temperatures of the core component \citep{1982ApJ...252..102C}, which represents the innermost section of the jet that is synchrotron self-absorbed at the given frequency, e.g., \cite{2011A&A...532A..38S}:
\begin{linenomath}
\begin{equation}
T_\mathrm{b} = 1.22 \times 10^{12} (1 + z) \frac{S}{\theta^2 \nu^2} \, \text{K},
\end{equation}
\end{linenomath}
where $S$ is the core flux density in Jy, $\nu$ the observing frequency in GHz, and~$\theta$ the full width at the half-maximum (FWHM) diameter of the fitted circular Gaussian component in mas. Among~the three available epochs, only in one case the core size exceeded the minimum resolvable size of the interferometer~\cite{2005AJ....130.2473K}. Consequently, we give a conservative lower limit to the brightness temperature $T_{\text{b}} \gtrsim 3.6 \times 10^{11}~\text{K}$ (Table~\ref{tab_X}). 

We use this lower limit to constrain the value of the Doppler-boosting factor:
\begin{linenomath}
\begin{equation}
\delta = \frac{T_\text{b}}{T_{\text{b,int}}}, 
\label{eq:Tb}
\end{equation}
\end{linenomath}
where $T_{\text{b,int}} = 4.1 \times 10^{10} \, \text{K}$ is assumed as the typical intrinsic brightness temperature measured for non-flaring quasars~\cite{2021ApJ...923...67H}, close to the equipartition value of \mbox{$T_{\text{b,eq}} \approx 5 \times 10^{10} \, \text{K}$ \cite{1994ApJ...426...51R}}. By~substituting $T_{\text{b}}$ into Equation~(\ref{eq:Tb}), we obtain the lower limit for the Doppler factor $\delta \gtrsim 8.8$.

Using the Doppler factor and the measured apparent jet component speed, we can determine the Lorentz factor ($\Gamma$) characterizing the bulk motion of the plasma and~estimate the inclination angle ($i$) of the jet relative to the line of sight~\cite{1995PASP..107..803U}. For~these calculations, we consider J5, the~innermost component with measured apparent speed at $8$~GHz. Its location is the closest to the VLBI core; therefore, it is the best to represent the innermost segment of the jet~\cite{2011A&A...532A..38S}. The~following equations apply, e.g., \cite{1995PASP..107..803U,2024Univ...10...97F}:
\begin{linenomath}
\begin{equation}
\Gamma = \frac{\beta_{\text{app}}^2 + \delta^2 + 1}{2\delta},
\end{equation}
\end{linenomath}
\begin{linenomath}
\begin{equation}
i = \arctan\left(\frac{2\beta_{\text{app}}}{\beta_{\text{app}}^2 + \delta^2 - 1}\right).
\end{equation}
\end{linenomath}
By substituting $\beta_\mathrm{J5}$ and the lower limit to $\delta$, we obtain $\Gamma \gtrsim 5.7$ and $i \lesssim 5.4^\circ$. The~calculated lower and upper limits are shown in Table~\ref{tab_par}.

Since the core brightness temperature lower limit measured in the latest epoch at 8~GHz ($T_\mathrm{b} > 12.6 \times 10^{11}$~K, Table~\ref{tab_X}) is nearly an order of magnitude higher than the conservative limit we used for calculating the jet parameters, we repeated the same process and obtained $\Gamma \gtrsim 15.8$ and $i \lesssim 0.6^\circ$ (Table~\ref{tab_par}). However, the~Doppler factor is likely overestimated here, because~the source may have undergone an outburst when its intrinsic brightness temperature could have exceeded the equipartition value~\cite{2021ApJ...923...67H}. The~increase in the sum of component flux densities (nearly $50\%$ compared to the previous year, Table~\ref{tab_X}) and the appearance of the new component J6 (Figure~\ref{fig:propermotion}) provide evidence for such an outburst. Therefore, we retain our conservative parameter~estimates.
\vspace{-4pt}
\begin{table}[H] 
\caption{Lower and upper limits for the inner jet~parameters.\label{tab_par}}

\begin{tabularx}{\textwidth}{CCCCC}
\toprule
\textbf{Epoch}	& \boldmath{$T_\mathrm{b}$} &
\boldmath{$\delta$}&
\boldmath{$\Gamma$} &
\boldmath{$i$} 
\\

\textbf{(year)}	& \textbf{(10}\boldmath{$^{11}$} \textbf{K)}&
\textbf{}&
\textbf{}&
\textbf{(°)}\\
\midrule
1994.612 &>3.6& $\gtrsim$8.8 & $\gtrsim$5.7 &$\lesssim$5.4 \\
2018.346& >12.6& $\gtrsim$30.8& $\gtrsim$15.8 & $\lesssim$0.6\\
\bottomrule
\end{tabularx}

\noindent{\footnotesize{Notes: Col.~1: mean observing epoch; Col.~2: brightness temperature of the core; Col.~3: Doppler factor; C\mbox{ol.~4: bul}k Lorentz factor; Col.~5: inclination angle.}}
\end{table}
\unskip

\section{Discussion}
\label{sec:Discussion}

For the most recent VLBI study of $z\geq3$ quasar jet proper motions~\cite{2025Univ...11...91G}, the~objects were drawn from a carefully defined initial sample of 102 high-redshift quasars~\cite{2021MNRAS.508.2798S}. Apart from requiring $z\geq3$, the~other primary selection criteria were the total flux density at $1.4$~GHz exceeding $100$~mJy, and~the declination between $-35^{\circ}$ and $+49^{\circ}$. The~subject of our study, the~quasar J1429+5406, also has high $1.4$ GHz flux density, but~its declination falls just out of the range considered in~\cite{2025Univ...11...91G}. Moreover, some of its redshift measurements available in the literature, ranging from $2.905$ \cite{2009ApJS..180...67R} to $3.036$ \cite{2017ApJS..233...25A}, are slightly lower than $3$. Nevertheless, it is physically similar to the AGN studied in~\cite{2025Univ...11...91G}, and therefore, its derived properties can be compared to those of the sample~objects.

Concerning the angular extent of its radio jet structure on VLBI scales, nearly $40$~mas, J1429+5406 is among the larger jetted quasars at $z\geq3$, although~we know a couple of more extended objects as well, e.g., \cite{2001ApJ...547..714L,2008A&A...489..517Y,2020A&A...643L..12S}. Based on our brightness distribution models (Table~\ref{tab_S}), the~outer steep-spectrum components J1 and J2 seen in the lower-frequency images (Figure~\ref{fig:images}) exhibit no detectable proper motion ($\lesssim 0.05$~mas\,year$^{-1}$) in the observed period (Figure~\ref{fig:propermotion_L}). This could be a combined effect of the relatively poorer angular resolution at $2.3$~GHz, the~insufficient time coverage, and~the diffuse nature of the emission features. A~possible physical cause for the apparent lack of detectable proper motion in these outer jet components is that the emission is not Doppler-boosted, either because the jet material slows down, e.g., in a standing shock~\cite{1988ApJ...334..539D} and/or it is inclined at a different angle to the line of sight than the inner ($\lesssim$10~mas) section.

The remarkable jet bending as seen projected onto the sky, at~$\sim$$20$~mas from the core (Figure~\ref{fig:images}), could be caused by a sudden reorientation of the jet due to its interaction with a dense clump of the interstellar medium. Future polarization-sensitive VLBI imaging could provide evidence for this scenario~\cite{2020NatCo..11..143A}. Alternatively, apparent changes in the jet direction could be due to small intrinsic variations in the spatial orientation of the jet, for~example, because~of precession detected in high-redshift sources as well~\cite{2016Galax...4...10R}. This could be caused by orbital motion in a binary black hole system inspiralling at the jet base, e.g., \cite{2014MNRAS.445.1370K}. In blazars, the~small jet inclination angle causes amplification of the bending when projected onto the sky. Similar cases are not uncommon at $z\ge3$. In~a study of a large sample, a~significant fraction ($6\%$) of the jets was found to have bending with more than $90^{\circ}$ \cite{2025Univ...11...91G}. 

Unlike the outer features of J1429+5406, the three inner jet components (J3, J4, and~J5; Table~\ref{tab_X}) exhibit detectable proper motion when modeled at $8$ and $5$~GHz, as~shown in Figure~\ref{fig:propermotion}. The~apparent proper motion of J3 was actually measurable at both $2.3$ and $8$~GHz, yielding similar values. 
Notably, the~apparent jet component speeds increase with increasing distance from the core in this inner part. Apparent jet acceleration on mas scales is common in AGN and can be either due to geometric of physical effects: a change in the jet direction with respect to the line of sight, or~an increase in the intrinsic jet speed (or the bulk Lorentz factor), probably the latter being the more common effect~\cite{2009ApJ...706.1253H}. According to VLBI measurements, jet flow acceleration can take place up to $\sim$$100$~pc deprojected distances~\cite{2015ApJ...798..134H}. Jet acceleration on pc scales can be attributed to, e.g.,~magnetic driving~\cite{2004ApJ...605..656V} or mass loading of the magnetohydrodynamic flow~\cite{2017MNRAS.469.3840N}. Thermal acceleration can also play a role by converting internal energy to bulk kinetic energy~\cite{2024A&A...683A.235R}.
Our VLBI data are rather poorly sampled in time, and therefore, we could only attempt determining linear proper motions for each individual component. Finding evidence for component acceleration, curved trajectories, and~jet precession in J1429+5406 would require long-term VLBI monitoring with denser time sampling, as~conducted for bright lower-redshift blazars~\cite{2021ApJ...923...30L}. Monitoring high-redshift sources needs more time since the intrinsic changes appear $(1+z)$ times slower in the observer's frame due to the cosmological time~dilation.

The structure and proper motion of J1429+5406 have earlier been studied with VLBI at $5$~GHz in the CJF (Caltech--Jodrell Bank Flat-spectrum) survey, albeit with a very time-limited set of data. The~object was part of the sample of 293 flat-spectrum sources~\cite{2008A&A...484..119B}. For~J1429+5406, the~apparent proper motion of their component designated with C1 (which corresponds to our J5 component) was estimated based on model fitting to VLBI data obtained at three epochs between $1993$ and $1998$. The~value $\mu_{C1}= (0.01\pm0.1)$~mas\,year$^{-1}$ has a formal error much larger than the measurement itself, and therefore, the apparent motion of C1 could not be conclusively determined during that short $5$-year period of time~\cite{2008A&A...484..119B}.
In contrast, our study spanning $24$ years of VLBI imaging at $8$~GHz yielded a significant apparent speed measurement for J5, $\beta_\mathrm{J5}=(4.7\pm 0.6)\,c$. 
The apparent speeds of quasar jets at high redshifts are typically in the range $(0.2-10)\,c$ \cite{2025Univ...11...91G}. In~comparison, $\beta_\mathrm{J5}\simeq5\,c$ is close to the average. On~the other hand, our component J3 with larger $\beta_\mathrm{J3}=(17.0\pm1.9)\,c$ is among the fastest-moving ones known in $z\ge3$ quasars~\cite{2020SciBu..65..525Z,2022ApJ...937...19Z}.

For further analysis of the geometric and physical parameters of the inner jet, we considered the J5 component. This is the closest to the core with  detections in all three available epochs at $8$~GHz, allowing us to estimate its apparent speed. Using the parameters of the core (i.e., the~jet base), we were able to establish a lower limit to its brightness temperature, $T_{\text{b}} \gtrsim 3.6 \times 10^{11}~\text{K}$. This value is nearly an order of magnitude higher than the equipartition brightness temperature~\cite{1994ApJ...426...51R}, confirming that the radio emission of the jet is Doppler-boosted. The~core components in the high-redshift sample of~\cite{2025Univ...11...91G} typically show high ($\gtrsim$10$^{10}$\,K) brightness temperatures. Our estimated inclination angle ($i \lesssim 5.4^{\circ}$) confirms that J1429+5406 is a blazar, as~the value is well below $10^{\circ}$ \cite{2022Galax..10...35P}. The~Lorentz factor ($\Gamma \gtrsim 5.7$) is consistent with the Lorentz factors described in two papers~\cite{2022ApJ...937...19Z,2025Univ...11...91G} on major samples of high-redshift sources, which present values between approximately 1 and 32, while their median is $\sim$$6$.

The blazar J1429+5406 has also been studied in the $\gamma$-ray and soft X-ray bands~\cite{2020ApJ...897..177P}. It was part of a sample of $142$ X-ray-detected AGN with $z\geq3$, consisting of 9 confirmed blazars and 133 blazar candidates. Here, the source was classified as a $\gamma$-ray-detected blazar. According to this high-energy study~\cite{2020ApJ...897..177P}, the~full sample was characterised by mean magnetic field strength $\overline{B}=(1.0 \pm 0.5)$\,G, mean Lorentz factor $\overline{\Gamma}= 7.0 \pm 1.9$, and~mean Doppler factor $\overline{\delta}= 12.3 \pm 2.3$. For~J1429+5406 in particular, the~values obtained are the following: magnetic field strength $B=2.4$\,G, Lorentz factor $\Gamma_1=11$, and~Doppler factor $\delta_1=16.5$. Our calculated lower limit of the Doppler factor ($\delta \gtrsim 8.8$) is consistent with the value determined from broadband spectral energy distribution (SED) modeling~\cite{2020ApJ...897..177P}. The~same applies to the Lorentz factor, with~the VLBI measurements providing a lower limit of $\Gamma \gtrsim 5.7$. It is interesting to note that while SED modeling suggests $\overline{\delta}\simeq 12$ for the Doppler factors~\cite{2020ApJ...897..177P}, VLBI observations of $z\geq3$ sources~\cite{2025Univ...11...91G} indicate that Doppler factors exceeding $10$ are~rare. 

Based on the limits we derived for its jet parameters, J1429+5406 appears similar to the typical known high-redshift radio-loud AGN~\cite{2024Univ...10...97F}. However, we note that we followed a conservative approach when selecting the $8$ GHz brightness temperature measurement that is compatible with all three epochs analyzed (Table~\ref{tab_X}). This led to the lower limits $\delta \gtrsim 8.8$ and $\Gamma \gtrsim 5.7$. However, the~Doppler factor could be much higher if we consider $T_\mathrm{b} > 1.26 \times 10^{12}$\,K obtained from the latest measurement in $2018$ (Table~\ref{tab_X}). Consequently, the~Lorentz factor might in fact significantly exceed our conservative lower limit, reaching at least $\Gamma \approx 15.8$, and~the jet inclination can also be somewhat smaller ($i \approx 0.6^{\circ}$). The~apparent compactness of the core, together with the uncertainties in deriving jet parameters from the currently limited 8 GHz VLBI monitoring data, highlights the importance of additional, more sensitive VLBI observations in the future. Such observations would enable tighter constraints on the core size and yield more accurate measurements of the jet component's proper motions in the blazar J1429+5406.

\section{Summary and~Conclusions}
\label{sec:Summary and Conclusions}

We analyzed archival VLBI imaging data of the high-redshift ($z=3.015$) quasar J1429+5406, taken at five different frequency bands, spanning about a quarter of a century. At~lower frequencies, components of the relativistic jet can be traced up to about $40$~mas from the radio core, with~a sharp turn occurring at $\sim$$20$~mas (Figure~\ref{fig:images}). This could be due either to a deflection of the jet by a cloud of dense interstellar matter or to the amplification, in~projection onto the sky, of~an intrinsically small change in the jet~direction.

Where data from multiple epochs are available, brightness distribution modeling allowed us to follow changes in the mas-scale radio structure of the source. The~apparent speeds characterizing the outward motion of the inner ($\lesssim$10~mas) jet components (Figure~\ref{fig:propermotion}) increase by their distance from the core. One of them (J3) reaches $\beta_\mathrm{J3}=(17.0\pm1.9)\,c$, which is among the highest values measured in $z \geq 3$ quasars to date~\cite{2020SciBu..65..525Z,2025Univ...11...91G}.

The radio emission of the source is Doppler-boosted. Using the apparent speed of the J5 component that is the closest to the jet base and the brightness temperature of the core measured at $8$~GHz, we arrived at conservative lower limits of the Doppler factor ($\delta \gtrsim 8.8$) and the bulk Lorentz factor ($\Gamma \gtrsim 5.7$). We could constrain the jet inclination angle to $i \lesssim 5.4^\circ$. The~latter value firmly places J1429+5406 among blazars which have jets oriented very close to the line of sight. The blazar nature of this source is further supported by its detection in $\gamma$-rays~\cite{2020ApJ...897..177P}.

The Doppler and Lorentz factors might be considerably higher than the lower limits estimated from the currently available data, potentially making J1429+5406 an outlier among high-redshift jetted quasars. To~confirm this with confidence, additional sensitive, high-resolution VLBI imaging at high observing frequencies ($\nu \ge 8$~GHz) would be~required.

\vspace{6pt} 

\authorcontributions{Conceptualization, S.F.; methodology, D.K. and S.F.; formal analysis, D.K.; writing---original draft preparation, D.K.; writing---review and editing, S.F.; visualization, D.K.; supervision, S.F. All authors have read and agreed to the published version of the~manuscript.}

\funding{This research was funded by the Hungarian National Research, Development, and Innovation Office (NKFIH), grant number OTKA K134213, and~by the NKFIH excellence grant TKP2021-NKTA-64.}

\dataavailability{The calibrated VLBI data are available from either the Astrogeo database (\url{https://astrogeo.org/cgi-bin/imdb_get_source.csh?source=J1429\%2B5406}, accessed on 18 April 2025) or, in~the case of the EVN data, from~the corresponding author upon reasonable request. The~raw EVN data are available from the EVN Data Archive (\url{http://archive.jive.nl/scripts/portal.php}, accessed on 18 April 2025) under project code EF022.} 

\acknowledgments{The EVN is a joint facility of independent European, African, Asian, and~North American radio astronomy institutes. Scientific results from data presented in this publication are derived from the following EVN project code: EF022. The~National Radio Astronomy Observatory is a facility of the National Science Foundation operated under cooperative agreement by Associated Universities, Inc. We gratefully acknowledge the use of archival calibrated VLBI data from the Astrogeo Center database~\cite{2025ApJS..276...38P} maintained by Leonid Petrov. D.K. is grateful for the support received from the observatory assistant programme of the Konkoly Observatory~\cite{konkoly}.}

\conflictsofinterest{The authors declare no conflicts of~interest.} 


\abbreviations{Abbreviations}{
The following abbreviations are used in this manuscript:\\

\noindent 
\begin{tabular}{@{}ll}
 AGN & Active galactic nuclei\\
 CJF & Caltech--Jodrell Bank Flat-spectrum (survey)\\
 EVN &  European VLBI Network\\
 FIRST & Faint Images of the Radio Sky at Twenty-centimeters (survey)\\
 FSRQ & Flat-spectrum radio quasar\\
 FWHM & Full width at half maximum\\
 mas & Milliarcsecond\\
 ROSAT & Roentgen Satellite\\
 SED & Spectral energy distribution\\
 SMBH & Supermassive black hole\\
 VLA & Karl G. Jansky Very Large Array\\
 VLASS & VLA Sky Survey \\
 VLBA &  Very Long Baseline Array \\
 VLBI & Very long baseline interferometry\\
\end{tabular}
}

\appendixtitles{no} 
\appendixstart
\appendix
\section[\appendixname~\thesection]{}

Tables~\ref{tab_L}--\ref{tab_U} show the parameters of the Gaussian brightness distribution model components fitted to the calibrated VLBI visibility data of J1429+5406 at $1.7$, $2.3$, $5$, $8$, and~$15.4$~GHz, respectively. Where the fitted component sizes were smaller than the minimum resolvable angular size of the interferometer~\cite{2005AJ....130.2473K}, the~latter value is given as an upper limit. The~uncertainties of the model parameters are estimated based on~\cite{2008AJ....136..159L}.
\vspace{-4pt}
\begin{table}[H] 
\caption{Parameters of the circular Gaussian components fitted to the $1.7$ GHz~data.\label{tab_L}}

\begin{tabularx}{\textwidth}{>{\centering\arraybackslash}X >{\centering\arraybackslash}p{1cm} >{\centering\arraybackslash}X
>{\centering\arraybackslash}X >{\centering\arraybackslash}X >{\centering\arraybackslash}X}
\toprule
\textbf{Epoch}	& \textbf{Comp.}	& \boldmath{$S$} &
\boldmath{$\theta$}&
\boldmath{$R$} &
\boldmath{$\phi$} 
\\

\textbf{(year)}	& \textbf{}	& \textbf{(mJy)}&
\textbf{(mas)}&
\textbf{(mas)}&
\textbf{(°)}\\

\midrule

\multirow[m]{4}{*}{2010.434} & J1& $89\pm19$& $17.7\pm1.0$ & $34.9\pm0.5$ &$93.6\pm0.8$\\

& J2 & $43\pm19$ & $5.0\pm0.3$& $23.4\pm0.1$& $132.8\pm0.3$\\

& J3& $117\pm20$& $5.1\pm0.3$ & $7.4\pm0.1$ &$133.9\pm1.1$\\

& C & $383\pm28$ & $1.9\pm0.1$ & 0 & \dots\\

\bottomrule
\end{tabularx}

\noindent{\footnotesize{Notes: Col.~1: mean observing epoch; Col.~2: component identifier; Col.~3: flux density; Col.~4: FWHM diameter; Col.~5: angular distance from the core (C); Col.~6: position angle of the component with respect to the core, measured from north through east.}}
\end{table}
\vspace{-14pt}

\begin{table}[H] 
\caption{Parameters of the circular Gaussian components fitted to the $2.3$ GHz~data.\label{tab_S}}

\begin{tabularx}{\textwidth}{CCCCCC}
\toprule
\textbf{Epoch}	& \textbf{Comp.}	& \boldmath{$S$} &
\boldmath{$\theta$}&
\boldmath{$R$} &
\boldmath{$\phi$} 
\\

\textbf{(year)}	& \textbf{}	& \textbf{(mJy)}&
\textbf{(mas)}&
\textbf{(mas)}&
\textbf{(°)}\\
\midrule
\multirow[m]{4}{*}{1994.612} & J1 &$62\pm33$ & $10.8\pm0.7$& $37.6\pm0.3$ & $90.4\pm0.5$ \\
& J2 &  $86\pm34$& $11.9\pm0.8$& $20.9\pm0.4$&$134.0\pm1.0$ \\
& J3 &  $60\pm33$& $1.6\pm0.1$& $5.75\pm 0.05$  &$146.5\pm0.5$ \\
& C &  $511\pm46$& $0.62\pm0.04$& 0 &\dots \\
\midrule
\multirow[m]{5}{*}{2017.235} & J1& $41\pm15$& $12.2\pm0.7$ & $37.8\pm0.4$ &$91.1\pm0.5$\\
& J2 & $69\pm16$ & $13.6\pm0.8$& $20.0\pm0.4$& $130.3\pm1.1$\\
& J3 & $32\pm15$ & $3.4\pm0.2$& $8.0\pm0.1$& $148.7\pm0.7$\\
& J5 & $90\pm16$ & $1.47\pm0.09$& $1.84\pm0.04$ & $131.0\pm1.3$\\
& C & $211\pm20$ & $0.79\pm0.05$& 0 & \dots\\
\midrule
\multirow[m]{5}{*}{2018.346}& J1  & $90\pm22$& $23.0\pm 1.5$ & $36.9\pm0.7$ & $91.0\pm1.2$
\\
& J2 & $63\pm22$ &$12.5\pm0.8$ & $21.4\pm0.4$ & $135.5\pm1.1$ \\
& J3 & $41\pm22$ &$4.5\pm0.3$ & $9.0\pm0.1$ & $145.8\pm0.9$ \\
& J5 & $133\pm23$ &$1.04\pm0.07$ & $2.00\pm0.03$ & $135.5\pm1.0$ \\
& C & $242\pm27$ &<0.25 & 0 & \dots \\
\bottomrule
\end{tabularx}

\noindent{\footnotesize{Notes: Col.~1: mean observing epoch; Col.~2: component identifier; Col.~3: flux density; Col.~4: FWHM diameter; Col.~5: angular distance from the core (C); Col.~6: position angle of the component with respect to the core, measured from north through east.}}
\end{table}
\unskip

\begin{table}[H] 
\caption{Parameters of the circular Gaussian components fitted to the $5$ GHz~data.\label{tab_C}}

\begin{tabularx}{\textwidth}{>{\centering\arraybackslash}X >{\centering\arraybackslash}p{1cm} >{\centering\arraybackslash}X
>{\centering\arraybackslash}X >{\centering\arraybackslash}X >{\centering\arraybackslash}X}
\toprule
\textbf{Epoch}	& \textbf{Comp.}	& \boldmath{$S$} &
\boldmath{$\theta$}&
\boldmath{$R$} &
\boldmath{$\phi$} 
\\

\textbf{(year)}	& \textbf{}	& \textbf{(mJy)}&
\textbf{(mas)}&
\textbf{(mas)}&
\textbf{(°)}\\

\midrule

\multirow[m]{5}{*}{1998.122} & J2 &$32\pm16$ & $12.0\pm0.5$& $20.1\pm0.2$ & $133.4\pm0.7$ \\

& J3 &  $24\pm16$& $4.3\pm0.2$& $6.21\pm0.09$&$139.5\pm0.8$ \\

& J4 &  $15\pm16$& <0.081& $2.172\pm 0.002$  &$130.14\pm0.04$ \\

& J5 &  $135\pm17$& $0.45\pm0.02$& $1.168\pm 0.009$  &$135.2\pm0.4$ \\

& C &  $319\pm20$& $0.32\pm0.01$& 0 &\dots \\
\midrule
\multirow[m]{3}{*}{2010.401} & J3& $74\pm13$& $10.3\pm0.6$ & $8.4\pm0.3$ &$132.8\pm2.1$\\

& J5 & $117\pm15$ & $0.95\pm0.06$& $1.24\pm0.03$& $135.6\pm1.3$\\

& C & $199\pm18$ & $0.20\pm0.01$ & 0 & \dots\\

\bottomrule
\end{tabularx}

\noindent{\footnotesize{Notes: Col.~1: mean observing epoch; Col.~2: component identifier; Col.~3: flux density; Col.~4: FWHM diameter; Col.~5: angular distance from the core (C); Col.~6: position angle of the component with respect to the core, measured from north through east.}}
\end{table}
\unskip

\begin{table}[H]
\caption{Parameters of the circular Gaussian components fitted to the $8$ GHz data and the calculated core brightness~temperatures.\label{tab_X}}

	\begin{adjustwidth}{-\extralength}{0cm}
		\begin{tabularx}{\fulllength}{CCCCCCC}
			\toprule
			\textbf{Epoch}	& \textbf{Comp.}	& \boldmath{$S$} &
\boldmath{$\theta$}&
\boldmath{$R$} &
\boldmath{$\phi$} &
\boldmath{$T_\mathrm{b}$}
\\

\textbf{(year)}	& \textbf{}	& \textbf{(mJy)}&
\textbf{(mas)}&
\textbf{(mas)}&
\textbf{(°)}&
\textbf{(10}\boldmath{$^{11}$} \textbf{K)}\\

			\midrule
\multirow[m]{4}{*}{1994.612} & J3 &$33\pm29$ & $6.1\pm0.5$& $5.2\pm0.2$ & $136.1\pm2.5$& \\

& J4 &  $68\pm30$& $0.54\pm0.04$& $1.67\pm0.02$&$137.9\pm0.7$ &\\

& J5 &  $90\pm30$& <0.22& $0.535\pm 0.008$  &$131.0\pm0.9$& \\

& C &  $259\pm35$& <0.212& 0 &...& >3.6 \\
                   \midrule
\multirow[m]{4}{*}{2017.235} & J3& $14\pm12$& $5.7\pm0.3$ & $8.7\pm0.1$ &$137.7\pm1.0$& \\

& J4 & $20\pm12$ & $1.72\pm0.09$& $3.26\pm0.04$& $132.4\pm0.8$& \\

& J5 & $55\pm12$ & $0.89\pm0.05$& $1.35\pm0.02$& $134.8\pm1.0$& \\

& C & $237\pm17$ & $0.179\pm0.009$& 0 & ...& $4.8\pm0.7$\\

\bottomrule
\end{tabularx}

\end{adjustwidth}
\end{table}

\begin{table}[H]\ContinuedFloat

\caption{{\em Cont.}}

\label{tab_X}

	\begin{adjustwidth}{-\extralength}{0cm}
		\begin{tabularx}{\fulllength}{CCCCCCC}
			\toprule
			\textbf{Epoch}	& \textbf{Comp.}	& \boldmath{$S$} &
\boldmath{$\theta$}&
\boldmath{$R$} &
\boldmath{$\phi$} &
\boldmath{$T_\mathrm{b}$}
\\

\textbf{(year)}	& \textbf{}	& \textbf{(mJy)}&
\textbf{(mas)}&
\textbf{(mas)}&
\textbf{(°)}&
\textbf{(10}\boldmath{$^{11}$} \textbf{K)}\\

			\midrule
\multirow[m]{5}{*}{2018.346}& J3  & $15\pm18$& $4.5\pm 0.2$ & $9.2\pm0.1$ & $142.8\pm0.6$& 
\\

& J4 & $21\pm18$ &$1.91\pm0.08$ & $3.35\pm0.04$ & $131.5\pm0.7$ &\\

& J5 & $48\pm18$ &$0.61\pm0.03$ & $1.67\pm0.01$ & $134.4\pm0.4$ &\\

& J6 & $113\pm18$ &$0.32\pm0.01$ & $0.323\pm0.007$ & $130.8\pm1.2$ &\\

& C & $283\pm21$ & <0.118 & 0 & ...& >12.6 \\
			\bottomrule
		\end{tabularx}
	\end{adjustwidth}
	\noindent{\footnotesize{Notes: Col.~1: mean observing epoch; Col.~2: component identifier; Col.~3: flux density; Col.~4: FWHM diameter; Col.~5: angular distance from the core (C); Col.~6: position angle of the component with respect to the core, measured from north through east; Col.~7: brightness temperature of the core.}}
\end{table}
\vspace{-14pt}

\begin{table}[H]
\caption{Parameters of the circular Gaussian components fitted to the $15.4$ GHz data and the calculated core brightness~temperature.\label{tab_U}}

	\begin{adjustwidth}{-\extralength}{0cm}
		\begin{tabularx}{\fulllength}{CCCCCCC}
			\toprule
			\textbf{Epoch}	& \textbf{Comp.}	& \boldmath{$S$} &
\boldmath{$\theta$}&
\boldmath{$R$} &
\boldmath{$\phi$} &
\boldmath{$T_\mathrm{b}$}
\\

\textbf{(year)}	& \textbf{}	& \textbf{(mJy)}&
\textbf{(mas)}&
\textbf{(mas)}&
\textbf{(°)}&
\textbf{(10}\boldmath{$^{11}$} \textbf{K)}\\

			\midrule
\multirow[m]{3}{*}{2010.149} & J5& $40\pm13$& $0.69\pm0.04$ & $1.57\pm0.02$ &$136.2\pm0.8$&\\

& J6 & $25\pm13$ & <0.041& $0.403\pm0.001$& $139.6\pm0.2$&\\

& C & $214\pm18$ & $0.152\pm0.009$ & 0 & \dots& $1.9\pm0.3$\\

			\bottomrule
		\end{tabularx}
	\end{adjustwidth}
	\noindent{\footnotesize{Notes: Col.~1: mean observing epoch; Col.~2: component identifier; Col.~3: flux density; Col.~4: FWHM diameter; Col.~5: angular distance from the core (C); Col.~6: position angle of the component with respect to the core, measured from north through east; Col.~7: brightness temperature of the core.}}
\end{table}

\begin{adjustwidth}{-\extralength}{0cm}

\reftitle{References}


\PublishersNote{}
\end{adjustwidth}
\end{document}